\documentclass[letterpaper, 10 pt, conference]{ieeeconf}
\IEEEoverridecommandlockouts
\usepackage[noadjust]{cite}
\usepackage{nicefrac}

\usepackage{ulem}
\usepackage{changes}
\usepackage{xcolor}
\usepackage{multirow, lipsum}
\usepackage{enumerate}
\usepackage{booktabs} 
\providecommand{\keywords}[1]
{	
  \textbf{\textit{Keywords---}} #1
}
\usepackage{amsmath}
\usepackage{physics}
\usepackage{hyperref}
\usepackage{url}
\usepackage{graphicx}
\usepackage{siunitx}
\usepackage{import}
\usepackage{placeins}
\usepackage{amsthm}
\usepackage{amssymb}
\usepackage{empheq}
\usepackage{ulem}
\usepackage{algorithm}
\usepackage{float}
\usepackage{multicol}
\usepackage{cleveref}
\usepackage[english]{babel}
\usepackage{bbm}
 
\theoremstyle{definition}

\newcommand{\RemoveForCamera}[1]{}

\newcommand{\RemoveCycle}[1]{}

\newtheorem{lemma}{Lemma}

\usepackage{algorithm}
\usepackage{algpseudocode}
\usepackage{amsfonts}
\usepackage{algorithm,algpseudocode}
\usepackage{dsfont}
\usepackage{bbm}
\usepackage{amsmath}
\usepackage{bm}
\newtheorem{thm}{{\bf Theorem}}
\newtheorem{corl}{Corollary}
\newtheorem*{remark}{Remarks}
\usepackage{amsmath,etoolbox}
\AfterEndEnvironment{equation}{\restarthintedrel}
\AfterEndEnvironment{align}{\restarthintedrel}
\newcounter{hints}
\renewcommand{\thehints}{\roman{hints}}
\newcommand{\hintedrel}[2][]{%
  \stepcounter{hints}%
  \if\relax\detokenize{#1}\relax\else\csxdef{hint@#1}{\thehints}\fi
  \mathrel{\overset{\textrm{(\thehints)}}{\vphantom{\le}{#2}}}%
}
\newcommand{\restarthintedrel}{\setcounter{hints}{0}}

\hypersetup{
    colorlinks=true,
    linkcolor=blue,
    filecolor=magenta,      
    urlcolor=cyan,
}

 \newcommand{\FocalNEAndOthers}[1]{}    
\newcommand{\hide}[1]{}
\newcommand{\N}{\mathcal{N}}

\newcommand{\eop}{\hfill{$\square$}}

\newcommand{\G}{{\cal G}}
\newcommand{\C}{{\cal C}}
\renewcommand{\P}{{\cal P}}
\newcommand{\M}{{\cal M}}
\newcommand{\D}{{\cal D}}
\newcommand{\B}{{\cal B}}
\newcommand{\A}{{\cal A}}

\newcommand{\BR}{{\cal {BR}}}
\newcommand{\bpsi}{{\bm \psi}}
\usepackage{bm}
\newcommand{\bphi}{{\bm \phi}}

\newtheorem{defn}{\bf Definition}
\newtheorem{propsn}{\bf {Proposition}} 

\newcommand{\sG}{{\mbox{\fontsize{5.2}{5.2}\selectfont{$\G$}}}}
\newcommand{\sC}{{\mbox{\fontsize{5.2}{5.2}\selectfont{$\C$}}}}

\newcommand{\sA}{{\mbox{\fontsize{4.7}{5}\selectfont{$\A$}}}}
\usepackage{bm}

\usepackage{algorithm}
\usepackage{algpseudocode}
\usepackage{amsfonts}
\usepackage{algorithm,algpseudocode}
\usepackage{dsfont}
\usepackage{bbm}
\usepackage{amsmath}
\usepackage{bm}
\newcommand{\sP}{{\mbox{\fontsize{5.2}{5.2}\selectfont{$\P$}}}}

\newcommand{\arxiv}[2]{#2}  

\usepackage{empheq}

\usepackage{textcomp}
\def\BibTeX{{\rm B\kern-.05em{\sc i\kern-.025em b}\kern-.08em
    T\kern-.1667em\lower.7ex\hbox{E}\kern-.125emX}}
\providecommand{\keywords}[1]
{
  \small	
  \textbf{\textit{Keywords---}} #1
}
\title{Cooperate or Compete: Coalition Formation in Congestion Games }
\author{Riya Sultana and Veeraruna Kavitha %
\thanks{R. Sultana, Veeraruna Kavitha  are with Industrial Engineering and Operations Research, IIT Bombay, Powai, Mumbai, 400076, India
{\tt\small \{riya{\_}sultana, vkavitha\}@iitb.ac.in}}%
}   
    
\begin{document}
\maketitle
\thispagestyle{empty}
\pagestyle{empty}
\begin{abstract}
 This paper investigates the potential benefits of cooperation in scenarios where  finitely many agents compete for shared
resources, leading to congestion and thereby reduced rewards. By appropriate coordination the
 members of the cooperating group (a.k.a.,  coalition) can
 minimize the congestion losses due to inmates, 
 while   efficiently facing the
competition from outsiders (coalitions indulge in a non-cooperative congestion game). 
The quest in  this paper is to identify the stable partition of coalitions that are not challenged by a new coalition. 

In contrast to the traditional cooperative games, the worth of a coalition in our game also depends upon the arrangement of the opponents. Every  arrangement leads to a partition and a corresponding
congestion game;  the resultant Nash equilibria (NEs) dictate the `worth'. The    analysis is further complicated due to the presence of multiple NEs for each such game. 


The major findings are: a) the grand coalition of all players is stable only in certain scenarios; b) 
interestingly, in more realistic scenarios, the grand coalition is not stable,  but other partitions (mostly the ones
with only one non-singleton group) are stable 
   or none of the partitions are stable.  When none  are stable,  there is a possibility of cyclic behaviour. Basically the players cycle through different collaborative arrangements constantly in pursuit of better selfish outcomes. 
  In essence, this research discovers  the stable partitions when the resources are congestible and  the coordination efforts carry a price.



\end{abstract}

\hide{

\section{Some results} 

In this section, 
We attempt to understand the results when the rest of the s are kept fixed and when the $\mu_1$ is increased starting from $\mu_2.$ 

\subsection*{Identical smaller links ($2\mu_N > \mu_2$)}  Here we assume smaller links $a_2, \cdots a_N $ are nearly identical; in other words, $\mu_N$ is significantly large as $\mu_N \approx \mu_2 $.  In this case 
\begin{itemize}
    \item One can observe severe congestion case, i.e., the case with $\nicefrac{\mu_1}{2}  < \mu_N$ at the start, i.e., when $\mu_1$ is close to $\mu_2$. Under severe congestion we proved theoretically the following results,
    
    \begin{itemize}
         \item  all partitions are stable at $\beta = 0$.
         \item We also have nestedness by  Theorem \ref{thm_nestedness} --- $\P_k$  is stable implies $\P_{k-1}$ is  stable.
         \item  The stability of partition like $\P_k=\{\C_{k}, \{1\}, \cdots \{1\}\}$ is implied by the stability of partition type $\tilde{\P}_k=\{\C_k, \C_{N-k}\}$ but not vice versa (see Theorem \ref{thm_nestedness} using bully NE).
         \item Considering pessimal NE, The per head utility of partition type $\P_k$ at $\beta=0$ strictly increases as $k$ decreases, i.e, $Z^0_k \le Z^0_{k-1}$ for all $2 < k\le N-1$ if $\frac{\mu_1}{2}<\mu_{N}$.

    \end{itemize}
\item We can observe near severe congestion case when, $\mu_N<\frac{\mu_1}{2}<\bar{\mu}$. Under this case we have results by simulation,
\begin{itemize}
    \item The per head utility of $\tilde{\P}_k$ is higher than the per head utility of $\P_k$, i.e., $\tilde{Z}^0_k\ge Z_k$.
    \item Partition $\P_k$ is not stable at $\beta=0$ but $\tilde{\P}_k$ is stable at $\beta=0$ when $\tilde{Z}^0_k=\sum_{i=1}^N\mu_i$.
\end{itemize}

\item We Also we have considered a case when $\frac{\mu_1}{2}>\bar{\mu}$; there is one major link and others are nearly identical. 
    \begin{itemize}
    \item When the difference $\frac{\mu}{2}-\bar{\mu} $ is slightly higher than $0$  then due to continuity of near severe some partitions are stable for small range and as the difference increases the stability region decreases for the partitions other than GC and for GC the stability region increases. And after a threshold of the difference no partition is stable for any value of $\beta$.  We proved theoretically that near major link regime (i.e., $\mu_1 >  N \mu_2$) $Z_k$ is increasing in $k$.  
    \end{itemize}
\end{itemize}

\subsection*{Asymmetric smaller links $2 \mu_N < \mu_2$} 
Here there is considerable difference b/w strengths  of smaller s, in particular $2 \mu_N < \mu_2 $ 
\begin{itemize}

\item One can not observe severe congestion case in this regime. 

\item However in a regime close to severe congestion (i.e., when $\mu_N < \mu_1/2 < {\bar \mu}$)
\begin{itemize}
    \item Further when $\mu_N < \mu_1/2 < {\bar \mu} - \mu_N$, GC is not stable for any $\beta$ by Theorem \ref{gc_nev_stable}.
    \item  As $k$ decreases range of stability of $\P_k$ increases. And $\tilde{\P}_k$ is not stable for any value of $\beta$ for all $k$.
    
\end{itemize}
\item Similar to identical links, we have defined near severe congestion case where, $\bar{\mu}-\mu_N < \frac{\mu_1}{2}<\bar{\mu}$.
\begin{itemize}
    \item For this case GC is stable but for small range. 
    \item  Similarly to the previous case As $k$ decreases range of stability of $\P_k$ increases and $\tilde{\P}_k$ is not stable for any value of $\beta$ for all $k$.
\end{itemize}
\item We Also we have considered a case when $\frac{\mu_1}{2}>\bar{\mu}$; there is one major link and some others are nearly identical and some others are much smaller than others.
\begin{itemize}
    \item When the difference $\frac{\mu}{2}-\bar{\mu} $ is slightly higher than $0$  then due to continuity of near severe the partition type $\P_k$ is more stable than partition type $\tilde{\P}_k$.
    \item Then for some range of  the difference $\tilde{\P}_k$ more stable than $\P_k$  , then above some threshold other than GC no partition is stable.
    \item As the difference increases the stability region of GC also increases (not monotonously).
    \item Similar to the identical links, in thus case also we proved theoretically that near major link regime (i.e., $\mu_1 >  N \mu_2$) $Z_k$ is increasing in $k$.
\end{itemize}
\end{itemize}
{\bf In all the numerical case studies, } we observe that GC is stable when $\beta $ is less than some threshold.}

\vspace{-2mm}
\section{Introduction }

Classical non-cooperative game theory often assumes agents compete for better utility independently.
 However, in many real-world scenarios,   
collaboration offers potential benefits (e.g., mergers like Vodafone and Idea in India, Walmart and JD.com in China, etc.). This paper studies such    possibilities in atomic congestion games. In these games,      multiple agents compete for shared-resources, where the congestion levels on various resources dictate the  rewards of individual users. 
The main premise of this paper is to investigate if some (or all)  agents considering joint decisions  can efficiently shape the congestion levels on the resources leading to a system  that is beneficial to the collaborating partners. On the other hand, the cost involved in coordinating and making joint decision  reduces the benefits of cooperation, thereby negatively influencing the collaboration efforts.

 Cooperative game theory provides tools to precisely analyze such scenarios where selfish players aim to derive better utilities for themselves, by making joint  choices   with a suitable subset of other players. 
The coalition formation in our atomic congestion game can be classified  as partition form game (\cite{paper2},\added{\cite{paper6}})  where the worth of any coalition also depends upon the arrangement of the opponents in the partition. Any partition can be  blocked by a new coalition if the latter anticipates  to achieve better; a partition is stable if any such  threat is empty. The new coalition needs to predict its worth based on the anticipation of   the possible  (potential
retaliatory)  rearrangement of the opponents --- 
%
%
%
%
%
 %
the worth of the new coalition  is dictated by the equilibrium of 
 the non-cooperative game among the coalitions formed after the rearrangement(s). 
%
 The analysis is  further complicated due to the existence of   
 multiple NEs in such games and the anticipation has to be robust also against all possible NEs in the new rearrangement(s).


 There is a substantial literature that investigates   the potential of cooperation in the non-atomic congestion games (where individual agents make negligible impact). Due to non-atomic nature, their focus is on the benefits of cooperation in terms of the improvements of the utilities of various subsets of agents (\cite{paper13,paper14}).  However similar questions take a very different flavour in the  case of atomic congestion games. 
 The focus here would be on the identification of the stable partitions (if any),  the size and the structure of stable partitions, the impact of communication cost on the stability of a given partition, etc.  
%
%
%
 In \cite{paper11} authors consider cooperative game theoretic study of atomic congestion games, however they  focus only on the stability of  grand coalition  and   do not consider the cost of communication.


We identify different categories of   congestion games, based on relative strengths of the shared resources and derive the stability properties of each category. Some properties are derived for the special case with equi-divisible congestion.

Consider a scenario where one of the resources offers significantly larger gain  --- multiple agents simultaneously sharing  the `big' resource derive better utilities than utilizing any other resource. In such cases   the grand coalition (GC) is stable, whenever the communication cost is not too large.  No other partition is stable for any value of the communication cost. Basically the players in GC 
utilize all the resources and amicably (time) share the total utility, while in other partitions all the groups compete to grab the big resource.

 In a  more realistic scenario, where more than one resource  offers  comparable gains, we again have two sub-categories.

 When the number of resources offering non-negligible gains is smaller than the number of players (we refer this as limited-resources scenario),    none of the partitions are stable,  if the communication costs are small. 
 \RemoveCycle{Agents can exhibit certain cyclic best response dynamics where players keep attempting to switch to a new configuration  (infinitely), in a bid to derive better  outcome.}%
 Surprisingly  
higher communication costs can induce stability: 
  some  partitions other than GC can  become stable at higher values of communication cost. However,   none of the partitions are stable when the communication costs are significantly high. 

In the second sub-category, where  the resources are neither too big nor too small, all the partitions are stable for small communication costs. 
However for higher communication costs,  the partitions with only one non-singleton group (not equal to grand coalition)  are stable.

Thus departing from the usual results reported in majority of literature that investigates the stability of grand coalition, our work illustrates that it is the other partitions that are rather stable; this is true under more realistic conditions on parameters. This contrast is  primarily due to more realistic modeling of the cooperative and non-cooperative aspects of the routing   problem using partition form games. 
Such a contrast is reported in the recent past, in the  context of many other applications,  by modeling them as  appropriate partition form games -- in lossy queuing systems that compete over market size, only duopolies are reported to be stable (\cite{paper2}), in small supply chains catering to essential items vertical cooperation between one supplier and a manufacturer is found to be stable under certain conditions (\cite{paper3}),  and for online auctions the authors in \cite{paper6} found that none of the agents prefer to bid together when the number of agents is great than four. 

\hide{ 
Conclusively the grand coalition is stable mostly in some unrealistic scenarios. In more realistic scenarios other partitions (mostly ones with one large coalition) can become stable or none of the partitions are stable.

into different varieties and derived a common analysis for each category.  

When multiple agents have to solve  a congested routing game (with one source and destination), some of them may find it beneficial to consider correlated decisions; however such decisions bring in a cost of coordination.  The resulting coalition formation game has partition form nature, where the worth of any coalition also depends upon the arrangement of the opponents.  This game is more complicated due to existence of multiple Nash Equilibria (NE) in the strategic form non-cooperative game between   the coalitions of the given partition.  
Any partition (and the operating NE) can be  blocked by a new coalition if the later anticipates to achieve better; a partition is stable if any such  threat is empty.  %

 }

\section{Problem description and Preliminaries}
We consider a routing problem from a common source to a common destination. Each user has infinite number of jobs to be shipped and  there are multiple paths or links connected from the  source to  the destination. In each time slot,  each user ships one job  after choosing a link.
%
Let $\mathcal{N}$ be the set of players and $\M$ the set of links, with   $|\N|=N$ and $|\M|=M$. 
If two or more users choose the same link in the same slot there would be collisions, which can result in delayed transmissions or equivalently smaller rewards. The  aim of the users is to derive maximum time averaged  reward over several time slots, which we capture via expected reward\footnote{
\label{footnote_one}
{\bf TU assumption:}
By virtue of the utilities being defined via time-averaged rewards, one can easily observe that \textit{Transferable Utility (TU) assumption} is valid for this setting -- for example,  two players can share the total utility $(\mu_1+\mu_2)$  derived by using two unequal links in  proportion  $(\xi \mu_1+(1-\xi)\mu_2)$ and $((1-\xi)\mu_1+\xi\mu_2)$ respectively,  when user 1 chooses link $a_1$, $\xi$ fraction of times and link $a_2$, $(1-\xi)$ fraction of times and user 2 chooses vice versa. Any  ratio of division can be achieved by choosing suitable $\xi.$ } in view of the assumed and expected ergodicity.

Users can  make coalitions/groups with an aim to make joint decisions (e.g., as in Footnote \ref{footnote_one}) after sufficient amount of communication  -- the aim here is to make joint (technically correlated) decisions that significantly improve the combined utility of the group. \textit{Basically the members of a coalition $\C$ can  minimize the losses of congestion due to other members of $\C$ by appropriate coordination and efficiently face the competition from outsiders.} Observe here that we have a transferable utility (TU) game (see Footnote \ref{footnote_one}). Formally, a partition $\P$  of $n$ coalitions is represented by: 
\begin{equation}
 \mathcal{P}=\{\C_1, \C_2,...,\C_n \}\text{  where,} \\\
 \cup_{i=1}^n{\C_i} = \mathcal{N} , \C_i\cap \C_j = \emptyset. \label{partition_ref}
 \end{equation}

             For each coalition $\C$, a central controller (one of the players) makes decision for the entire group; the decision is in terms of the set of links (one link per each member of $\C$) represented as strategy $\bphi_\sC$; it is important to observe here that two or more players can use the same link (simultaneously) if they find it beneficial. Let \textit{$\bphi = ( \bphi_\sC : \C \in \P ) $ represents the strategy profile of all the coalitions or equivalently that of all the players}. 
 The  instantaneous (slot-wise)  \textit{reward for any link decreases as the number of users pulling  that link  (in the given slot) increases}. The utility of a coalition $\C$ of  
  partition $\P$, when  the coalitions of $\P$  choose strategy profile $\bphi$, defined in terms of the expected reward, is given by:
\begin{eqnarray}
    U_{\sC}^{\sP}(\bphi)=\sum_{a \in \bphi_{\sC}}\gamma_a(\bphi_{\sC})\mu_a(\gamma_a(\bphi))-(|\C|-1)\beta ,\label{utility_def}
\end{eqnarray}
where, $\gamma_a(\bphi):=\sum_i 1_{\{ \phi_i =a\}}$ (see notations in section \ref{sec_notation})  is the number of times link $a$ has been chosen in $\bphi$,  $\mu_a(k)$ is the expected reward obtained from link $a$ when $k$   number of players simultaneously pull it, and $(|\C|-1)\beta$  is the total communication cost of coalition $\C$; here it is assumed that one player among the coalition communicates with all other members and hence the cost is $(|\C|-1)$ times the cost per communication $\beta$.
Without loss of generality (w.l.g.), we index  the links as below and consider `non-zero links':
\begin{eqnarray}
    \mu_{a_1}(k) \ge \mu_{a_2}(k)\ge \cdots \ge \mu_{a_M}(k) > 0 \mbox{ when  } k = 1 .\label{Eqn_monotone_utils_k_one}
\end{eqnarray}

 \subsection*{Worth of a coalition in a  partition}
From \eqref{utility_def},  the reward of a player/coalition for choosing a strategy depends upon the strategy of other players as well. \textit{This leads to a  non-cooperative congestion game among the coalitions in any partition $\P$, where the utilities are given by \eqref{utility_def}}. Then, the {\it worth} $\nu_\sC$  of any coalition $\C$  can be defined as the expected combined utility \eqref{utility_def} that the group derives under an appropriate NE of that game. It is clear that this worth depends upon the way the opponents choose their own coalitions, in other words,  it is more appropriate to represent the worth by $\nu_\sC^\sP$, explicitly indicating the dependence on $\P$.  Such games are a special class of coalition formation games called \textit{partition form games, where the worth of a coalition     depends  not only upon the members of the coalition but also upon the arrangement of the opponents} (see e.g., \cite{paper1,paper3}). 



We   assume that   the players know the link-mean functions $\{ \mu_a (k) \}$ and investigate the following question: which partition is stable, in the sense that, which groups of players find it beneficial to operate together. To answer this question, one first needs to define formally the notion of stability, in particular, for partition form games. This is discussed in  section \ref{sec_partition}.  Before proceeding towards that, we would conclude this section by making  an important  observation.  

We will observe in the later sections that the non-cooperative game between the coalitions, described   above, can result in multiple NEs.  Thus, in our problem, the worth of a coalition does not just depend upon the partition $\P$ (as in partition form games), it additionally depends upon the resulting NE. In other words, 
 the worth of any coalition $\C \in \P$ when the partition operates under an NE $\bphi^*  $ is more aptly represented by $\nu_{\sC}^{\sP}(\bphi^*)$. Further,  observe from \eqref{utility_def}: 
 \begin{eqnarray}
     \nu_{\sC}^{\sP}(\bphi^*) &=& \nu_{\sC}^{0,\sP}(\bphi^*) -(|\C|-1)\beta,  \mbox{ where, }  \nonumber\\
       \nu_{\sC}^{0,\sP}(\bphi^*) &=& \sum_{a \in \bphi_{\sC}^*}\gamma_a(\bphi_{\sC}^*)\mu_a(\gamma_a(\bphi^*)). \label{Eqn_nuC_P_Phi} 
 \end{eqnarray}
 Here $\nu_{\sC}^{0,\sP}(\bphi^*)$ represents the worth with zero communication cost; \textit{it is important to observe here that the set of NEs does not depend upon the communication cost, as this  is fixed component of \eqref{utility_def}, once $\P$ is given.} 
 Let the set of  NEs  corresponding  to  partition $\P$ be denoted by   $\Phi^{\sP}$. We represent an NE by $\bphi$ (after dropping   $*$, when there is clarity).

\section{Partition form Games and Stability Notions}
\label{sec_partition}

Given the worth of   coalition $\nu_\sC^\sP(\bphi)$ for each $\C \in \P$, the next notion is that of the pay-off vector $\bpsi$. Any pay-off vector $\bpsi = [\psi_1, \cdots, \psi_N]$  has $N$-components with component $\psi_k$ representing the payoff (or utility) of player $k$.   A pay-off vector $\bpsi$ (in TU-game) is   consistent with partition $\P$, if there exists  an NE   $\bphi \in \Phi^{\sP}$ such that (see \cite{paper7} for more details): 
\begin{eqnarray}
\sum_{j \in \mathcal{C}_i}\psi_j = \nu_{\sC_i}^{\sP}(\bphi),  \mbox{ for all } \C_i, \mbox{ and, } \psi_j \ge 0 \mbox{ for all } j \label{worth_coalition}.
\end{eqnarray}
Basically, the worth of the coalition $\nu_\sC^\sP(\bphi)$ has to be  divided completely among  its members, and this must be true for all $\C \in \P$.
Define the set of consistent pay-off vectors with regard to the underlying NE  $\bphi$ and partition $\P$ as,
\vspace{-2mm}

{\small \begin{align}
    \Psi_{\bphi}^{\sP} :=  \left \{\bpsi=[\psi_1,\cdots,\psi_N] \in \mathcal{R}_{+}^N:\sum_{j \in \mathcal{C}_i}\psi_j=\nu_{\sC_i}^{\sP}(\bphi) \mbox{ $\forall$ } i \right \}. \nonumber 
\end{align}}
We refer the tuple  $ (\P,\bphi,\bpsi)$ 
as  equilibrium-driven configuration if the pay-off vector
  $\bpsi$ is consistent with $(\P, \bphi)$,  and if $\bphi$ is an NE under $\P$, i.e., if $\bphi \in \Phi^{\sP}$ and $\bpsi \in \Psi^{\sP}_{\bphi}$. Note that an equilibrium-driven configuration specifies not only a partition and a payoff vector, but also   an NE  (this is in contrast to what is usually seen in partition form games \cite{paper3}, \cite{paper7}).

\subsection {Blocking by coalition and Stability}
\textit{A configuration $(\P,\bphi,\bpsi)$ is said to be blocked by a coalition $\C \notin \P$} if for any    partition $\mathcal{P}'$ containing $\C$ and any NE $\bphi' \in \Phi^{\sP'}$ there exists a pay-off vector $\bpsi' \in \Psi_{\bphi'}^{\sP'}$  such that the following is true:
\begin{eqnarray}
 \psi'_j > \psi_j \mbox{  for all }  j \in \mathcal{C}.
 \label{Eqn_block}   
\end{eqnarray}
Intuitively, a coalition  $\C \notin \P$ blocks a configuration  if the
members of $\mathcal{C}$ have an incentive to ‘break’ one or more coalitions of $\P$ to come together and form
a new coalition. In particular, it is possible to allocate payoffs within the blocking coalition $\mathcal{C}$
such that each member of $\mathcal{C}$ achieves a strictly greater payoff, irrespective of any (potentially
retaliatory) rearrangements among the agents outside $\mathcal{C}$ and the NE strategy profile that they may settle to. We \textit{refer this  as blocking via pessimal anticipation rule, extending similar notion existing in  the literature (\cite{paper1,paper3}) to the case with multiple NEs}.

 From \eqref{Eqn_block}, it is easy to verify that   the  configuration $(\P,\bphi, \bpsi)$ is blocked by a coalition  $\C$  under pessimal anticipation if,
    \begin{align}
        \sum_{j \in  \C}  \psi_j <\underline{\nu}_{\sC}
\mbox{  
where, }  
\underline{\nu} _\sC \ := \ \inf_{\P : \C \in \P }     \inf_{  \bphi \in  \Phi^{\sP} }   \nu_{\sC}^{\sP}({\bphi} ). \hspace{4mm}\label{blocking_cond}
\end{align}

Now we have few important definitions. 
\begin{defn}\label{def_stability}
    \textit{ (a) A configuration $(\P,\bphi, \bpsi)$  is said to be stable} if there does not exist a coalition $\C \notin \P$ that can block it. \\
    (b) \textit{A partition-NE (P-NE) pair $(\P,\bphi)$ is said to be stable} if there exists at least one configuration $(\P,\bphi, \bpsi)$ involving the pair, which   is stable. \\
   (c) \textit{In similar lines, a partition $\P$ is said to be stable} if, there exists at least one configuration $(\P,\bphi, \bpsi)$ involving the partition, which   is stable. 
\end{defn}
Consider \textit{`fair'} pay-off vector    $\bpsi_f$, under which the worth(s) are divided equally within the coalition(s), i.e.,    define
    $\psi_{f, j} = \nu_{\sC}^\sP(\bphi)  / | \C| $ for all  $j \in \C \in \P.$
We have the following result concerning the fair payoff vector:
\begin{lemma} \label{lemma_1_PNEpair}
  A P-NE pair $(\P,\bphi)$ is stable if and only if the configuration $(\P,\bphi, \bpsi_f)$ is  stable.   
\end{lemma}%
\noindent{\bf Proof:} \ Say the P-NE pair $(\P,\bphi)$  is stable, thus there exists a configuration $(\P,\bphi, \bpsi)$ which is not blocked by any $\C \notin \P$. 

Any   $\C\notin \P$ can in general have $q_i :=  | \C_i \cap \C |$ members from  coalition $\C_i$ with $0\le q_i \le |\C_i|$. One  can  as well assume that the $q_i$ players  (of coalition $\C_i$)  in $\C$ are deriving the  least pay-off   (if they don't find it beneficial to block then, any other combination of $q_i$ agents of $\C_i$ will also not find it beneficial to block) and  then clearly,
\begin{eqnarray}
    \sum_{j \in \C}\psi_{f,j} = 
\sum_{\C_i \in \P } \frac{q _i}{|\C_i|} \nu_{\sC_i}^\sP (\bphi) \ge \sum_{\C_i \in \P }  \sum_{j \in \C \cap \C_i} \psi_j  \ge   {\underline \nu}_\sC,\label{stability_byanticipating}
\end{eqnarray}
as $\C$ does not   block  the configuration $(\P,\bphi, \bpsi)$.
And this is true for any $\C \notin \P$, thus  the configuration $(\P,\bphi, \bpsi_f)$ is  stable. The converse is true by Definition \ref{def_stability}.
\eop

By  above lemma, we have identified a payoff vector that represents a stable configuration, in fact 
the stability is synonymous with 
  the fair pay-off vector.  
 Before proceeding further,  we   introduce some important notations.

\subsection{Notations}
\label{sec_notation}
  When all the players operate together the coalition is called grand coalition (GC) and is represented by $\G$. 
When  player $i$ operates alone, $\A_i := \{i\}$ represents the singleton coalition. A generic coalition 
is denoted by $\C_i$ or $\C$. 
The partition of all players  is called \textit{GC partition} and is denoted by $\P_{\sG}=\{\G\}$. When all the players operate alone we have  partition $\P_{\sA}=\{\A_1,\cdots,\A_N\}$, which we refer to as\textit{ ALC partition}.


The players are identical, hence    w.l.g.,   any partition   $\P = \{\C_1, \C_2  \cdots, \C_n\}$   is represented with the following convenient details: a) for each $i$,  let $l_i$ represent cardinality $|\C_i|$,  $\hat{l}_{i}:=\sum_{j=1}^{i-1}l_j$ (with  ${\hat l}_1 
   := 0$) represent the starting index of the coalition $i$;   and  then,  b) the members of   coalition  $\C_i$ of $\P$ can be indexed as below:
$$
   \C_i = \{ \hat{l}_{i}+1, \cdots, \hat{l}_{i}+l_i\}. $$ 
Further, 
w.l.g., we also assume  $ l_{i+1} \le l_{i}$ for each $i.$ 

Recall   any strategy profile of  coalition $\C_i$   is represented  by  $\bphi_{\sC_i} $ 
$ = \{ \phi_j: j \in \C_i\}$
and $\bphi = ( \bphi_{\sC_1}, \cdots, \bphi_{\sC_n})$ is   the strategy profile of all the players.  
With the new notations, 
on the other hand, 
 strategy profile  $\bphi = (\phi_1,   \cdots, \phi_N) $  under $\P$, implies the following sub-vector as the sub-strategy profile of    coalition $\C_i \in \P$:
\begin{eqnarray}
   \mbox{strategy  }   \ \bphi_{\sC_i}  = (\phi_{{\hat l}_{i}+1}, \cdots, \phi_{{\hat l}_{i}+l_i}),    \mbox{  for each } i . \label{Eqn_Notation}  
\end{eqnarray}
We represent $\mu_{a}(1)$ as $\mu_{a}$ for any $a\in \M$, { and   $\mu_{a_i}$ as $\mu_i$ for any $a_i\in \M$.
%
{\it The proofs of the theorems and lemmas are given in the Appendix \ref{appendix}.}

\subsection{GC Partition}
We would like conclude this section by drawing attention to  a special point related to GC. 
In GC partition all the players operate together, thus there is no game involved. Rather the players optimize the  joint utility and so  (see~\eqref{utility_def}): 
\begin{eqnarray}
\nu_\sG^{\sP_{\sG}}  :=   \max_\bphi \left \{ \sum_{a\in \M} \gamma_{a} (\bphi)\mu_{a} (\gamma_{a} (\bphi) ) \right \} - (N-1) \beta .\label{eqn_unq_optimizer}
\end{eqnarray}
\textit{We assume there exists a unique optimizer for the above;} basically, we avoid possibilities like   $2\mu_1(2) = \mu_{N}+ \mu_1   $ (with $M\ge N$) to get uniqueness  (see \eqref{Eqn_monotone_utils_k_one}).  We now have the following result which is easy to verify:  
\begin{lemma} \label{gc_uniq_ne}
 Assume  $\mu_1 (2) \le \mu_{N}$ or   consider equi-divisible congestion model,  where $\mu_a(k) = \mu_a/k$ for all $k, a$. Let $M\ge N$.
    Then the  unique optimizer of \eqref{eqn_unq_optimizer} is $\bphi_\sG = (a_1, \cdots, a_N)$ and $\nu_\sG^{\sP_{\sG}} = \sum_{i=1}^N\mu_i - (N-1)\beta$. \label{lemma_2_unq_optimizer}\eop
 %
  
\end{lemma}
Lastly, it is easy to observe that  the pessimal anticipated utility for grand coalition, i.e., for $\C=\G$, gets simplified; basically  ${\underline \nu}_\sG  =  \nu_\sG^{\sP_{\sG}} $.

\section{Influence of communication cost on stability}

The worths   $\{\nu_\sC^{\sP}(\bphi) \}$,  and the anticipated worths  $\{\underline \nu_\sC \}$,  at any per communication cost $\beta \ge 0$  are related to the corresponding ones at zero communication cost (specially represented by ${\underline \nu}_\sC^{0}$ etc.,) as below.   From \eqref{utility_def} and \eqref{Eqn_nuC_P_Phi}   (using simple computations) for any $\P, \C, \bphi$:
\begin{equation}
 {\nu}^\sP _\sC (\bphi)  =  {\nu}^{0,\sP} _\sC (\bphi)  - \left ( |\C|- 1\right )   \beta \mbox{ and }
  \underline{\nu} _\sC  = \underline{\nu} _\sC^0 - \left ( |\C|- 1\right )   \beta. \label{nu_not_anticip}
\end{equation}
 Using these relations, one can  derive 
 the conditions for stability of 
  any  P-NE pair $(\P, \bphi)$  at any  $\beta \ge 0$    in terms of the worths at zero communication cost, $\{\nu_\sC^{0,\sP}(\bphi) \}$ and $\{ {\underline \nu}_\sC^{0} \}$. 
In this section we derive these relations and then the stability results as a function of $\beta$.

 Recall  (see  proof of Lemma \ref{lemma_1_PNEpair}) that  any  $\C \notin \P$ 
can in general have $q_i :=  | \C_i \cap \C |$ members from  coalition $\C_i$ for each $i$  (with $0\le q_i \le l_i$) and that the  P-NE pair $(\P,\bphi)$  is stable if it is not blocked  by all such $\C$. Thus one can rewrite the conditions for stability as below (see \eqref{blocking_cond}):

\begin{propsn}\label{thm_stability_1}
Define $q_i (\C) =| \C_i \cap \C|$ for each $\C_i \in \P$ and  $\C \notin \P$.
Then the P-NE pair $(\P, \bphi)$ is stable if and only if the following conditions are satisfied for each $\C \notin \P$:\begin{eqnarray}
    \Gamma_\sC^\sP (\bphi)  + \D_\sC^\sP \beta  & \ge &  0,  \label{cond_stability} \mbox{ where, }  \D_\sC^\sP \ :=  \  \sum_{i=1}^n   \frac{q_i(\C)}{l_i} - 1    \nonumber \\
\label{Eqn_Stability_conditions}  
    \mbox{and }
\Gamma_\sC^\sP (\bphi) &:=& \sum_{i=1}^n \frac{q_i(\C)}{l_i} \nu_{\sC_i}^{0,\sP}(\bphi) - {\underline \nu}^{0}_\sC .  
\end{eqnarray}  
\end{propsn} 
\noindent
{\bf Proof:} By Lemma \ref{lemma_1_PNEpair}  stability of $(\P,\bphi, \bpsi_f)$ is equivalent to the stability of $(\P,\bphi)$.  
Now the result follows from 
   equations \eqref{Eqn_nuC_P_Phi}, \eqref{stability_byanticipating} and \eqref{nu_not_anticip}. \eop
   \hide{
Without loss of generality one can assume that the $q_i$ players  (of coalition $\C_i$) with least pay-off attempt to deviate, as they are the most vulnerable and if they don't find it beneficial to deviate any other combination of $q_i$ agents of $\C_i$ will also not find it beneficial to deviate. Then it is easy to verify that,
$$
\sum_{j \in \C \cap \C_i} \psi_j \le \frac{q _i}{l_i} \nu_{\sC_i}^\sP (\bphi).
$$
\eop}

Say, for some $\C$ and P-NE pair $(\P,\bphi)$ we have
 $\D_{\sC}^\sP < 0$ in \eqref{Eqn_Stability_conditions};  for example,  this is true for a split, i.e., when   $\C \subsetneq \C_i $ for some~$i $. For such cases, if  $\Gamma_{\sC}^\sP(\bphi) <  0$, then that particular $\C$ can block the P-NE pair for all $\beta \ge 0$.  This type of arguments can be used to derive the following result, using the  definitions  given below:
 %
%
%
 \begin{eqnarray}
  \B^{--} &=&  \B^{--} (\P) \ = \ \{\C \notin \P:\D_{\sC}^{\sP} \le 0, \Gamma_\sC^\sP(\bphi) < 0\},  \nonumber\\
    \B^{+-} &=&\{\C \notin \P:\D_{\sC}^{\sP} > 0,\Gamma_\sC^\sP(\bphi) < 0\}, \label{set_def}\\
    \B^{++}&=&\{\C \notin \P:\D_{\sC}^{\sP} \ge 0, \Gamma_\sC^\sP(\bphi) \ge 0\}, \nonumber\\
    \B^{-+}&=&\{\C \notin \P:\D_{\sC}^{\sP} < 0, \Gamma_\sC^\sP(\bphi) \ge 0\},  \mbox{ and } \nonumber\\ 
     \bar{\beta}_{u}^{\sP} (\bphi) &:=& \hspace{-3mm} \min_{\C \in \B^{-+}} \bar{\beta}_{\sC}^\sP(\bphi), \nonumber\ 
    \bar{\beta}_{d}^{\sP}(\bphi):=\max_{\C \in \B^{+-}} \bar{\beta}_{\sC}^\sP (\bphi), 
    \mbox{ with }  \nonumber  \\ 
    \bar{\beta}_{\sC}^\sP(\bphi) &=&-\frac{\Gamma_\sC^\sP(\bphi)}{\D_\sC^\sP}. \nonumber 
\end{eqnarray}
Here we utilize the standard convention that the maximum and minimum of an empty set are  $0$ and $\infty$ respectively.

The immediate observations are, 
\textit{  ${\bar \beta}_d^\sP(\bphi) , {\bar \beta}_u^\sP (\bphi)\ge 0$  and   $\{\C \notin \P\}= \B^{+-}\cup\B^{--}\cup \B^{++}\cup \B^{-+}$.} Using these, 
we  have the first important result (proofs are in Appendix \ref{appendix}):
 \begin{thm} \label{thm_stabililty}
 Consider any  P-NE pair ($\P,\bphi)$ and    the  definitions as in \eqref{set_def}. Then the following are true,\\
      i) if   $\B^{--} = \emptyset$ and  
      ${\bar \beta}_d^\sP(\bphi) \le   {\bar \beta}_u^\sP(\bphi)$,   the P-NE pair  is stable only for $\beta \in [{\bar \beta}_d^\sP(\bphi) , \  {\bar \beta}_u^\sP(\bphi)]$, \\
     ii) otherwise the P-NE pair is not stable for any $\beta \ge 0$, and \\
         iii)  if  $\B^{--}=\emptyset$ and if $\P \ne \P_{\sA}$ (the ALC partition), we have $\bar{\beta}_u^{\sP} (\bphi) <\infty$.  \eop
 \end{thm}

\noindent{\bf Remarks:} (1) One may anticipate the stability to be monotone in communication cost, however, surprisingly this may not always be the case ---   for the scenarios in   Theorem \ref{thm_stabililty}.(i), the said non-monotonicity is true further when $\bar{\beta}_d^\sP (\bphi) > 0$.

(2) Say a
  P-NE pair $(P,\bphi)$ is stable for some communication cost $\beta$ (thus does not fall under the scenarios covered  in Theorem \ref{thm_stabililty}.(ii)). Then the  pair is stable   for   an interval  of the communication costs by part (i). Further by part (iii),  this interval is finite for any collaborating partition (i.e., when $\P$ is not equal to ALC, $\P_\sA$). 
 
 (3) Some  of  the    P-NE pairs  may not be stable either with zero communication cost or when the  communication cost is large, but could be  stable for some intermediate values of~$\beta$ --- this is true when $\bar{\beta}_d^\sP(\bphi) > 0$ and when $\P \ne \P_\sA$. 

In section \ref{numerical_analysis}, we derive some numerical insights  directly using Definition \ref{def_stability};   
the numerical examples  of this section   reinforce 
  each of the above mentioned possibilities.  For now we concentrate on theoretical insights that can be derived using  
Definition   \ref{def_stability} and  Theorem  \ref{thm_stabililty}.  

By virtue of the fact that only splits can potentially block the GC,  and because there exists a unique P-NE pair  $(\P_{\sG},\bphi_\sG )$ involving GC (see Lemma \ref{gc_uniq_ne}),  
we have the first important corollary  of Theorem \ref{thm_stabililty}, when $M\ge N$.  
\begin{corl} \label{gc_stability} Assume $M \ge N$.
a) If the P-NE pair $(\P_\sG,\bphi_\sG)$  is not stable at $\beta = 0$, then the P-NE pair and hence  GC is not stable for any $\beta  \ge  0$.  \\
b) Otherwise, 
    there exists a threshold   $\bar{\beta}_{u}^{\sP_\sG} \in [0, \infty)$, such that  $(\P_{\sG},\bphi_\sG )$ and hence GC  is stable only for    $\beta \le \bar{\beta}_{u}^{\sP_\sG}$. \eop
\end{corl} 

In similar lines the 
ALC  partition (all players operate alone) is blocked only by mergers and hence the following:
\begin{corl} \label{ALC_stability}
There exists a threshold  $\bar{\beta}^{\sP_{\sA}}$
such that ALC partition is stable for all $\beta \ge \bar{\beta}^{\sP_{\sA}}$.  
\end{corl}
 {\bf Proof:} Here both the sets $\B^{--}$ and $\B^{-+}$ are empty. Thus $\bar{\beta}_u^{\sP_{\sA}}=\infty$. By Theorem \ref{thm_stabililty}, any P-NE pair  $(\P_{\sA},\bphi)$ involving ALC partition is stable for $\beta \ge \bar{\beta}_{d}^{\sP_{\sA} } (\bphi).$  The result follows with $\bar \beta^{\sP_\sA}  := \min_{\bphi, NE} \bar{\beta}_{d}^{\sP_{\sA} } (\bphi)$ by virtue of Definition \ref{def_stability}.  \eop  


\medskip

\noindent {\bf Remarks:} (1) Thus, there are  only two possibilities regarding the stability of grand coalition: a) either  GC is not stable irrespective of the communication cost; or b) GC is stable   if the  communication cost is below  the threshold $\bar{\beta}_{u}^{\sP_\sG}$. 

(2) On the other hand, ALC is stable once the communication cost is  above a threshold. 

(3) Further by Theorem \ref{thm_stabililty} and Definition \ref{def_stability}(c),  \textit{   for
 any partition $\P$ other than ALC   there exists a finite threshold, 
 $$
 \bar \beta_u^{\sP} := \max_{\bphi, NE} \bar \beta_u^{\sP} (\bphi),
 $$ 
 such that  the partition is not  stable when $\beta > \bar \beta_u^{\sP}$.} 

(4) To conclude,  
\textit{when per communication cost $\beta$ is sufficiently high, none of the partitions other than ALC }are  stable; i.e., the players do not prefer coordination when the cost of communication is sufficiently expensive.


\begin{figure*} [htbp]
\vspace{1cm}
\begin{minipage}{0.5\textwidth}
    \centering
    \vspace{1.4cm}
 \includegraphics[trim={0cm 0cm 0cm 4.3cm}, width= 6cm, height=3.8cm]{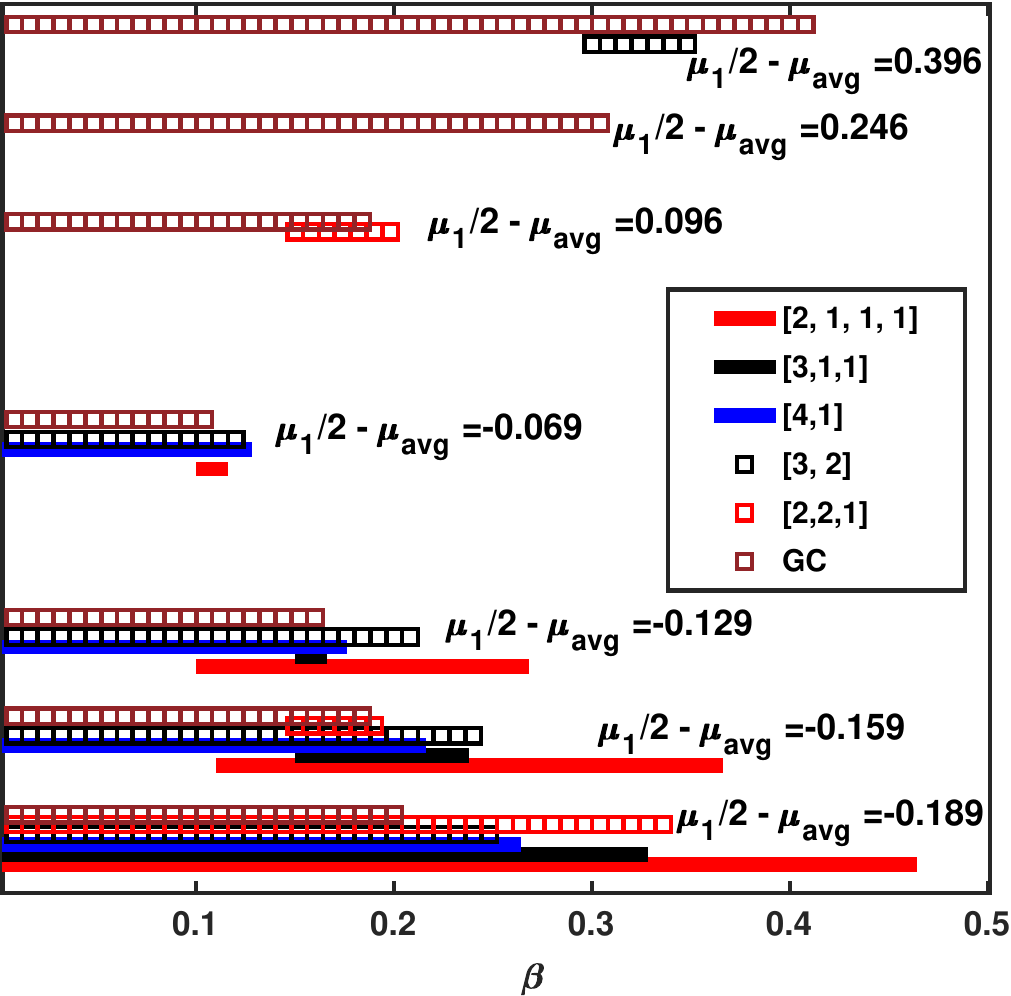}
    \caption{Stability Analysis: none of the   arm is negligible   }
    \label{fig:enter-label}       
\end{minipage}
\begin{minipage}{0.5\textwidth}
    \centering
    \vspace{1.5cm}
 \includegraphics[trim={0cm 0cm 0cm 4.3cm}, width= 6cm, height=3.78cm]{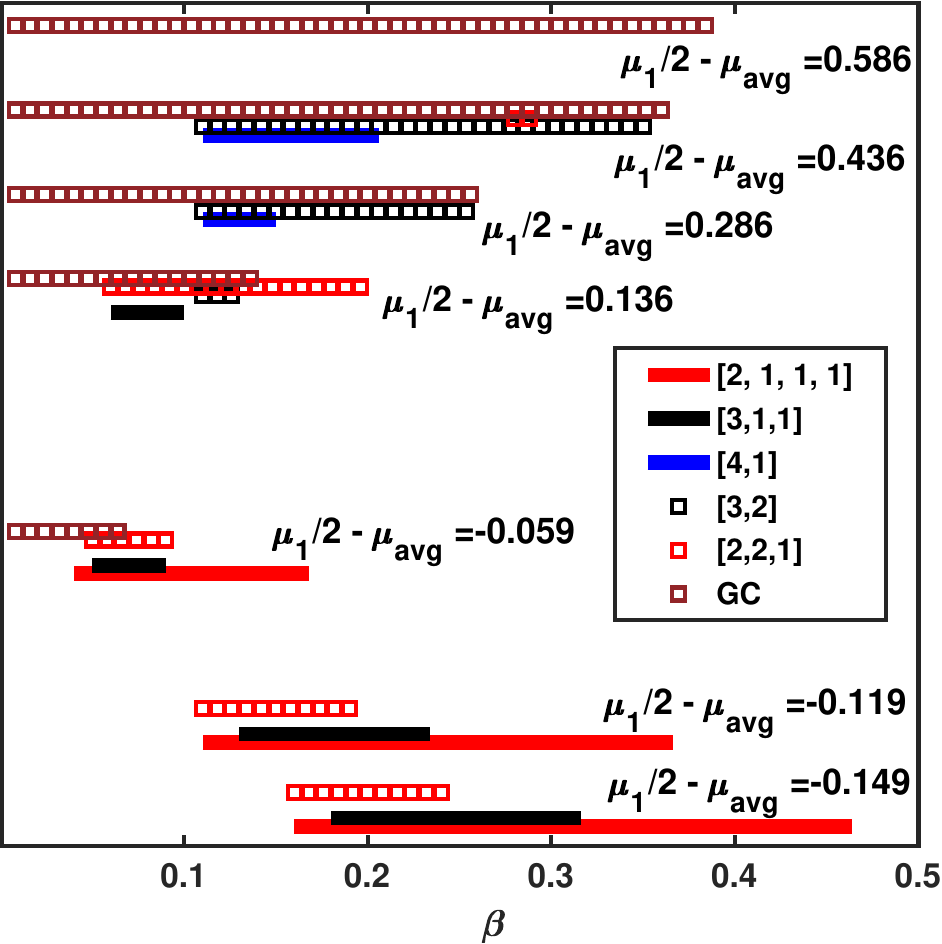}
    \caption{Stability Analysis  limited-resources, arm 5 is negligible }
    \label{fig:enter-label2}
 \end{minipage}
 \vspace{-4mm}
\end{figure*}
 \section{Stable Partitions}
We now investigate further   stability properties  and compare various  partitions. 
Towards this, we begin with a   scenario at which the links are severely congested, represented by the condition $\mu_1(2) < \mu_N$ (\textit{henceforth, we also assume $M \ge N$} or without loss of generality $M=N$); basically the players may find it   beneficial to use distinct links, as the utility with two or more players  choosing the same link is smaller even for $a_1$.

\subsection{Severe congestion} \label{severe_congestion}

Consider a system where  a player gets inferior utility upon choosing a link that is already   chosen by some  other  player modelled by   the following assumptions/conditions:
\begin{eqnarray}
\mu_1 (2) &<& \mu_N   \mbox{ and } \nonumber  \\
    \mu_{l}(k) &\ge& \mu_{l+1}(k)  \mbox { for all }   k  \le N ,   l < M  .
    \label{Eqn_monotone_utils_k_all} \label{eqn_assum}
\end{eqnarray}
\hide{
This type of case we refer as severe congestion. We have discussed the stability of partitions based on per communication cost $\beta$ and also compared the range of stability of different partitions considering focal NE. 

{\color{red}
{\bf Assumptions:}
Further, 
 we assume the following 
 throughout the report:
\begin{itemize}
    \item The number of links is greater than or equal to the number of players, i.e., $M \ge N$ 
    
\item Assume, 
\begin{eqnarray}
    \mu_{1}(k) \ge \mu_{2}(k)\ge \cdots \ge \mu_M(k), \forall k  \le N.
    \label{Eqn_monotone_utils_k_all} \label{eqn_assum}
\end{eqnarray} 
\end{itemize} }
 }
 For this case we have the following  (proof in Appendix \ref{appendix}):
%
\begin{thm} 
\label{thm_sev_ne}
  i)   Let $\bphi$ be any permutation of $\{a_1, \cdots, a_N\}$, the set of first $N$ links as in  \eqref{Eqn_Notation}. Then   $\bphi$ is an NE for    any partition. Further no other strategy profiles are   NE.   \\
  ii) All P-NE pairs and   all partitions are stable at $\beta=0$. \eop
\end{thm} 

Thus 
every partition is stable at $\beta = 0$ under \eqref{eqn_assum}.  
When a coalition attempts to deviate, 
the deviating group is unsure of the retaliation from the others and  more importantly is unsure of the NE that they can reach in the new configuration --- observe any permutation of $(a_1, \cdots, a_N)$ is an NE and thus a coalition of size $l$ can capture any $l$-subset (the worst $l$-subset is  $(a_{N-l+1}, \cdots, a_N$)) of the top $N$-links at the resultant NE. 
This is the precise reason for the stability of all P-NE pairs at $\beta = 0$ under    pessimal anticipation.

 Thus by Theorem \ref{gc_nev_stable}, any partition $\P$ is stable exactly for $\beta \in [0, {\bar \beta}_u^\sP]$, where ${\bar \beta}_u^\sP < \infty$ is  a $\P$ dependent constant.

\subsection {Non severe and equi-divisible congestion}

We  now consider  scenarios with   $\mu_1(2)>\mu_{N}$,  where     some group(s) of players may  find it beneficial to  use a common link (at the same time). As one may anticipate    the analysis of such systems is far more complicated. We thus consider a specific yet interesting  case study,  that with equi-divisible congestion: \vspace{-2mm}
$$
\mu_a( k) = \frac{\mu_a}{k} \mbox{ for any } a \le N \mbox{ and } k. 
$$
Basically,  the total payoff for any link gets divided  equally among all the agents simultaneously choosing the link.   \textit{Further, we consider the scenarios that have at least one (pure strategy) NE  for each partition\footnote{It is well known that the  unweighted congestion games are potential  games and hence  a pure strategy NE exists. But the same is not guaranteed for weighted  games (see e.g., \cite{paper8}). }.} 

With  $\mu_1(2)>\mu_{N}$, 
the players may find it beneficial to  repeat the `major link', 
   that is the link $a_1$ with larger mean. 
  However   the expected value $\mu_1$ of major link may or may  not be too large. In following we show two contrasting results based on the relative strength of $\mu_1$:  
  \begin{thm} \label{gc_nev_stable}
  \label{stability_gc}
 i)  GC is not stable for $\beta=0$ and hence also for any $\beta >0$, if 
    \begin{eqnarray}
        \mu_{N} < \frac{\mu_1}{2}< \bar{\mu}-\mu_{N}, \mbox{ where, }  \bar{\mu}:=\frac{\sum_{i=1}^N\mu_{i}}{N}. \label{int-cond} 
    \end{eqnarray}
    ii) Say  $\nicefrac{\mu_1}{2}>\mu_k$ for some $2\le k \le N$. Then any  partition $\P$   with size, $|\P|\ge k$ is blocked by GC and hence is not stable at $\beta=0$.\nonumber\\
    iii) Say  $\nicefrac{\mu_k}{2}>\mu_N$ for some $1\le k \le N-1$. Then any  partition $\P$   with size, $|\P|> N-k$ is blocked by GC and hence is not stable at $\beta=0$. \eop
\end{thm} 
\noindent{\bf Remarks:} (1) By part (i), when the major link is not too major,  the GC may not be stable even with zero communication cost. 

 (2)  On the other hand, if $\mu_1$ is significantly large (say $\nicefrac{\mu_1}{2}>\mu_2$ in part (ii)), then  GC  blocks all other partitions at $\beta = 0$. 

(3) The theorem  also suggests a  very  interesting possibility}   where  \textit{none of the partitions are stable at $\beta = 0$.}  One such  example is the case when     $  \nicefrac{\mu_1}{2}< \bar{\mu}-\mu_{N}$ and $\nicefrac{\mu_{N-1}}{2}>\mu_N$:  a) 
    from part (iii)   any  partition other than GC is not stable at $\beta=0$; and b) from part (i) GC is not stable. These conditions represent a very realistic scenario, which we  refer hence forth as `limited-resources' scenario:   here few of the links have negligible returns and the rest  are not significantly different. 
    
    (4) \textit{Thus in limited-resources scenarios (the major link $a_1$ is not too major and   $a_N$ has negligible utility) none of the partitions are stable at $\beta~=~0$}. Further GC is not stable for any communication cost $\beta$. However the  other partitions may become stable for larger $\beta$ (these possibilities are illustrated in section \ref{numerical_analysis}).

 \hide{ 

    We have proved in part (i) of Theorem \ref{gc_nev_stable} that GC will be blocked by a coalition of size $(N-1)$. And also in part (iii) we have shown that GC can block partition $\P=\{\C,\{1\}\}$ where $|\C|=N-1$. Thus there is a limit cycle among GC and partition $\P$.

  Consider that the above condition $\nicefrac{\mu_1}{2}>\mu_k$ is not satisfied and say $\mu_{k+1}<\nicefrac{\mu_1}{2}<\mu_k$, then consider a partition of size $|\P|=k$ where $\P=\{\C, \{1\} \cdots, \{1\}\}$. One can observe that, $|\C|=N-k+1$. Then consider a strategy profile $\bphi$ of $\P$ where, the $k-1$ independent players choose distinct links among $\{a_2 \cdots, a_{k}\}$. Then if $\nicefrac{\mu_2}{2}<\mu_N$ then the best response of $\C$ against the opponents is:
    \begin{eqnarray}
        \BR_\C(\{a_2, a_3 \cdots, a_k\})=\{a_1, a_{k+1}, a_{k+2} \cdots, a_N\}
    \end{eqnarray}
   (If not i.e., $a \in \BR_\C(\{a_2, a_3 \cdots, a_k\})$ such that $a \in \{a_2, a_3 \cdots, a_k\}$ then we have contradiction for $\bphi_{\C}$ being the best response against $\bphi_{-\C}$ as we have $\nicefrac{\mu_a}{2}<\mu_N$).
   And similarly the best response of independent players against opponents is among $\{a_2 \cdots, a_k\}$ (if not i.e., $a_1 \in \bphi_{\{1\}}$ then we have a contradiction having best response $\bphi_{\{1\}}$ against opponents as we have $\nicefrac{\mu_1}{2}<\mu_k$.) Thus, $\bphi$ is an NE. So, there exists a P-NE pair under these conditions such that  all players choose distinct links. Thus GC can't block it. We aren't arguing $\P$ is stable or not but we attempted to prove that GC can not block the partition.}

Towards  
  deriving further insights, we consider  numerical analysis in the immediate next.

\hide{
    By Remark \ref{remark_gc} and Theorem \ref{gc_stability}, GC is not stable for any $\beta \ge 0$ under the condition \eqref{int-cond}.
 
By simulation, we have observed that, if $\mu_{N-1} > \frac{\mu_1}{2}>\mu_{N}>\frac{\mu_2}{2}$ there exist a P-NE pair $(\P,\bphi)$ where, $\P=\{\C,\{i\}\}$ (where, $|\C|=N-1$ and $i$th player is independent) and $a_i \in \bphi$ for all $1\le i \le N$, i.e., all players choose distinct links in this NE, then $\P$ is stable for $\beta=0$.

Also we have derived the condition for blocking all other partitions  by GC using the relation among $\bar{\mu}$ and highest link mean $\mu_1$.

 By simulation, we have observed that under this condition, GC is stable for $\beta=0$.
}

%

\section{Numerical Analysis}\label{numerical_analysis}


We numerically compute  NEs of different partitions, worths  of coalitions under various P-NE pairs, the pessimal anticipated utilities and finally derive the stability  analysis for  a   given 
 set of  $\{\mu_i\}_{i\le N}$ and for  equi-divisible congestion. 
 We compute  the stability at  $\beta = 0$ and  also derive the range of per communication cost $\beta$ for which the given partition is stable.  
Our aim in this section is to understand relative stability  of various partitions across various $\beta$, as $\mu_1$  (the expected value of the major link) varies:    we fix $(\mu_2, \cdots, \mu_N)$  and increase $\mu_1$  starting from $\mu_2$.

 In the first case study  presented in  Figure \ref{fig:enter-label},  all the   smaller links are important (opposite to the limited-resources scenario) and   equal
 $ (\mu_2, \mu_3, \mu_4, \mu_5)= (.52,  .5,  .45, .3)$; also observe $\nicefrac{\mu_2}{2}<\mu_5$.

 Before we proceed, we would like to explain the figures. We have $N = 5$, hence have 6 distinct partitions (other than ALC) for each sub-case-study. So, in the figures each sub-case-study corresponds to one value of $\mu_1$ and is represented by a cluster of (at maximum) 6 horizontal bars. Each bar within the cluster corresponds to one partition (e.g., bar made up of brown squares corresponds to GC, while the blue thick bar corresponds to partition $[4, 1]$, etc.,). The  location of the bar corresponding to a particular partition represents the range of $\beta$ for which the said partition is stable. If a particular partition is not stable for any value of $\beta$ for some sub-case-study, then the corresponding bar will be missing in that cluster (e.g., in Figure \ref{fig:enter-label}, partition $[3, 1, 1]$, represented by black bar, is never stable for $\nicefrac{\mu_1}{2} - {\bar \mu} = -0.069$).

In  Figure \ref{fig:enter-label}
each cluster corresponds to a  distinct value of $\mu_1$, the value of $\nicefrac{\mu_1}{2} - {\bar \mu}$ is shown to the right of the cluster.   The bottom most cluster  
corresponds to the case with severe congestion. 
 For this cluster,  all partitions are stable at $\beta = 0$ as  proved in Theorem \ref{thm_sev_ne}. On the other hand, for the top 3 clusters, none of the  partitions  other than GC are stable at $\beta = 0$; this reaffirms Theorem~\ref{gc_nev_stable} (part ii)  with  $\nicefrac{\mu_1}{2}>\mu_2$. 

We now discuss other interesting observations based on this numerical study: a) most of the partitions are stable for some range of $\beta$ only when $\mu_1$ is relatively small;  b) \textit{when the major link becomes significantly superior (as in the top clusters)  most of the partitions are not stable  for any range of $\beta$};   c)  more interestingly \textit{the partitions with one non-singleton coalition are `more stable' than those with larger  number of non-singleton coalitions} --- partitions $[2, 1, 1, 1]$ or $[3, 1 , 1]$ or $[4,1]$ (red,  black or blue continuous bars)  
appear to be stable for a bigger set of parameters than the partitions $[2, 2, 1]$   or  $[3,2]$ (red or black squares); and d)  there are many scenarios (mostly with larger $\beta$) for which GC is not stable but other partitions are stable.


 A second   case study  that can include limited-resources regime is   in Figure \ref{fig:enter-label2} where the smaller  links are  significantly different:  we set  $ (\mu_2, \mu_3, \mu_4, \mu_5) = $   (.52, .5,.45, .1); here $\nicefrac{\mu_2}{2}>\mu_5$.    
 
For this example, one can not observe severe congestion. 
Many observations are similar to the experiment of Figure~\ref{fig:enter-label}.  However there \textit{are some striking differences}:  a)   \textit{there are no stable partitions near $\beta = 0$ (the bottom two clusters correspond to limited-resources regime discussed right after Theorem \ref{gc_nev_stable}}); b) more interestingly  \textit{a larger communication cost induces stability} (for e.g., in bottom most clusters some partitions become stable for $\beta>0.1$ and none before) ; and c)  the partitions with one non-singleton coalition, are `more  stable'  in the above cases with large communication costs.

\section*{Conclusions}
When multiple agents have to solve  a congested routing game (with one source and one destination), some of them may find it beneficial to consider correlated decisions; however, such decisions bring in a cost of coordination.  The resulting coalition formation game has partition form nature, where the worth of any coalition also depends on the arrangement of the opponents.  
The worth of any coalition in any partition is specified by the NE of the corresponding non-cooperative congestion game among the coalitions. 
The analysis is more complicated due to the existence of multiple NEs in each such congestion game.  
Any partition (and  an operating NE) can be  blocked by a new coalition if the later anticipates to achieve better; a partition is stable if any such  threat is empty. 

\hide{Also we have some interesting results is that considering focal NE stability of some partition is implied by the stability of some other partitions under severe congestion. And we have compared the stability of partitions based on communication cost. }

In the presence of a major link  with significantly   larger utility (than others), the grand coalition  is stable.  No other partition is stable for any value of the communication cost. 

However, in a  realistic scenario with limited-resources, where the  major link  is  comparable with some of the others, and where  few links are negligible,      
  none of the partitions are stable  with small communication costs.
Higher communication costs can induce stability: 
  some  partitions other than GC can  become stable.

When none of the links are major (gains of none of the links 
is significant
 enough that players find it beneficial to use them simultaneously)  all the partitions are stable at smaller communication costs. Further, partitions with one major coalition (and not equal to grand coalition) are stable at higher communication costs.  

Conclusively the grand coalition is stable mostly in some unrealistic scenarios. In more realistic scenarios other partitions (mostly the ones
with only one non-singleton group) can become stable or none of the partitions are stable.  

\arxiv{\bibliographystyle{IEEEtran}
\bibliography{references}
\bibliographystyle{alpha}
}{

}

\section{Appendix} \label{appendix}

\underline{\textbf{Proof of Theorem \ref{thm_stabililty}:}}
For notational simplicity in this proof we discard $\bphi$ from many terms. 
To begin with observe:
 $\{\C \notin \P\}=\B^{+-}\cup \B^{++}\cup \B^{-+} \cup \B^{--}$.
In all the proofs given below, 
we either negate or verify the   stability condition  \eqref{cond_stability}   and then the required  results follow using Proposition \ref{thm_stability_1}. \\ 
 \textbf{Proof of Part(i):} By definitions in  $\eqref{set_def}$,   for any $\C  \in  \B^{+-} $,   the stability condition \eqref{cond_stability} is satisfied for all 
 $\beta \ge \bar{\beta}_d^\sP.$ Thus such a $\C$ will not block $\P$ for those $\beta$. 
Similarly   the P-NE pair can not be blocked by any coalition $\C \in \B^{-+}$ if $\beta \le \bar{\beta}_{u}^\sP  $. 
And any $\C \in  \B^{++}$ can not block the P-NE pair for any $\beta$. 
This proves part (i) by
 Proposition \ref{thm_stability_1}.

 \noindent \textbf{Proof of Part(ii):}
 From \eqref{set_def}, if $\B^{--}\neq \emptyset$ then there exists a coalition $\C \notin \P$ such that, $\D_{\sC}^{\sP} \le 0$ and $\Gamma_\sC^\P(\bphi) < 0$. Then the stability condition   \eqref{cond_stability} is not satisfied with that particular $\C$  for any $\beta \ge 0$. 
Similarly if  $\bar{\beta}_{d}^\sP  >  \bar{\beta}_{u}^\sP  $ (see Part(i)), there exists at least one $\C$ either in $\B^{-+}$ or in $\B^{+-}$ which does not satisfy the stability condition   \eqref{cond_stability}  for every $\beta \ge 0$. 
 

{\bf Proof of Part(iii):}  If $\P = \P_{\sA}$ (the ALC partition), then only mergers can potentially block it, in other words for any  $\C \notin \P$, $\D_\sC^\sP >0$, then $\B^{-+} = \emptyset$.

For the given P-NE pair, if $\B^{--}=\emptyset$  then clearly,  $\B^{-+}\ne \emptyset$. Only minimum over non-empty set is finite and hence the proof. \eop

\underline{\textbf{Proof of Corollary \ref{gc_stability}:}}
Consider a P-NE pair ($\P_{\sG},\bphi_{\sG}$). Clearly $\D_{\sC}^{\sP_{\sG}}<0$ for all $\C \notin \P_{\sG}$ as only splits can block GC. So, $\{\C \notin \P_{\sG}\}=\B^{--} \cup {\B^{-+}}$.

If the P-NE pair is not stable for $\beta=0$, i.e., if $\Gamma_{\sC}^\P(\bphi)<0$, then $\B^{--}\ne \emptyset$. Then, by theorem \eqref{thm_stabililty}, the P-NE pair is not stable for any $\beta \ge 0$. 

Otherwise by \eqref{set_def} $\{\C \notin \P_{\sG}\}= \B^{-+}$. Thus by theorem \ref{thm_stabililty}, the P-NE pair is stable for $\beta$ in the range $[{\bar \beta}_d^\sP , {\bar \beta}_u^\sP]$ where, ${\bar \beta}_d^\sP=0$ and ${\bar \beta}_u^\sP<\infty$ by convention. Hence the proof.\eop

\underline{\textbf{Proof of Theorem \ref{thm_sev_ne}:}}
 Consider any partition, $\P = \{\C_1, \C_2,\cdots,\C_n\}$. 
Say there exists an NE $\bphi^*$ (notations  are as in section \ref{sec_notation}).  Our first claim   using \eqref{Eqn_monotone_utils_k_one} is  that    $\phi_i^* \in \{a_1, \cdots, a_N\}$ for each $1 \le i \le N$.   If not and say $\phi_i^* \in \{a_{N+1}, \cdots,  a_M\}$ for some $i$, then clearly   there exists an  $a \in \{a_1, \cdots, a_N\}$ such that $a \ne \phi_i^*$ for all $i \in \mathcal{N}$. Then  the utility of the corresponding coalition  (i.e., $\C$ such that $i \in \C$) can be   improved strictly (unilaterally) by replacing $\phi^*_i$ with  $a$ (see \eqref{Eqn_monotone_utils_k_all}), contradicting that $\bphi^*$ is an NE. Thus we can conclude that if there exists an NE $\bphi^*$ then, $\phi_i^* \in \{a_1, \cdots, a_N\}$ for all $1 \le i \le N$.

Consider any $\bphi$ as in hypothesis and any partition $\P$.  Let $\C = \C_i$ be one of its members and let $\bphi_{-\sC}$ represent the tuple of strategies  of opponent coalitions, $\{ \bphi_{\tilde \sC} :  {\tilde \C} \ne \C,  {\tilde \C}  \in  \P  \}$.
Then by $\eqref{Eqn_monotone_utils_k_one}$, the best response against $\bphi_{-\sC}$ is given by:
\begin{eqnarray*}
    \mathcal{BR}_{\{\sC\}}(\bphi_{-\sC}) &=& \{a_1,\cdots,a_N\}\setminus \ \bphi_{-\sC} ,
\end{eqnarray*}
as:  i) as in the first step, any  in  $\{a_{N+1}, \cdots, a_M\}$ can not be a part of the  optimal choice; and ii) by  hypothesis, it is optimal not to repeat any  (as for any $j \ne j'$, we have  $\mu_j(2) < \mu_{j'}$). Thus $\bphi$ is an NE for partition $\P.$
In fact using similar logic, no BR contains a strategy profile with an   repeating. This completes the proof of part (i).

\noindent{\bf Part(ii):} In view of   the above result, the pessimal utility \eqref{blocking_cond} of any coalition  $\C$ (say, $|\C|=k$) simplifies to:
\begin{eqnarray}
    \underline{\nu}_\C=\sum_{p=N-k+1}^N \mu_p-(k-1)\beta, \label{eqn_pessimal_sev}
\end{eqnarray}

Consider any P-NE pair $(\P,\bphi)$. Consider the configuration $(\P,\bphi,\bpsi_f)$ where $\bpsi_f$ is as in Lemma \ref{lemma_1_PNEpair}.
%
By  Theorem~\ref{thm_sev_ne}, 
  NE $\bphi$ is any permutation of $\{a_1, \cdots, a_N\}$. Hence
for any coalition $\C \notin \P$   the following holds:
 \begin{eqnarray}
     \sum_{i \in \C}\psi_{i,f} \ge  \sum_{p=N-|\C|+1}^N\mu_p=\underline{\nu}^0_{\sC}.
 \end{eqnarray}
Thus the P-NE pair $(\P,\bphi)$ is  stable by equation \eqref{blocking_cond} and Definition \ref{def_stability}.
\eop

\underline{\textbf{Proof of Theorem \ref{gc_nev_stable}:}}
{\bf Proof of Part (i):} We consider  $\beta = 0$ and show that GC is not stable, then by Theorem \ref{gc_stability} it is not stable for any $\beta > 0$.
For equi-divisible congestion it is easy to verify that, as in  Lemma \ref{gc_uniq_ne} $\bphi_\sG = (a_1, a_2, \cdots, a_N)$ uniquely optimizes the related objective function. 
\hide{

there exists an unique P-NE pair  $(\P_{\sG},\bphi_\sG )$ involving GC

Now it is immediate that optimization defining the worth of GC is solved by (a strategy with first $N$ distinct s):
\begin{eqnarray}
\nu^{\sP_{\sG}}=\nu_{\sG}^{\sP_{\sG}}(\bphi^*)= \max_{\bphi \in \Phi^{\sP_\sG}}U_{\sG}^{\sP_{\sG}}(\bphi^*)=\sum_{i=1}^N\mu_{i}. \label{2}
\end{eqnarray}}
Thus any configuration  involving $\P_{\sG}$ is of  the form, $(\P_{\sG},\bphi_\sG,\bpsi_\sG)$, where   $\bpsi_\sG$ is any payoff vector consistent with $(\P_\sG, \bphi_\sG)$. 
We will now show that the coalition  $\C = {\cal N} - \{i^*\}$ of $(N-1)$ players, with  $i^* \in \arg \max_j \psi_{\sG,j}$, i.e., the coalition formed by  leaving  out the player with maximum payoff in $\bpsi_\sG$,  will block $(\P_{\sG},\bphi_\sG,\bpsi_\sG)$ under the given hypothesis.  

Observe that the only partition possible after such a deviation is $\P  = \{ \C, \{i^*\} \}$  and hence  the pessimal anticipation coincides with pessimal across all possible NE  with $\P$.
By collective rationality ($\sum_i \psi_{\sG, i} = \nu_\sG^{\sP_\sG}$) and by definition of $i^*$ the total payoff derived by players of $\C$ under  $(\P_{\sG},\bphi_\sG,\bpsi_\sG)$ can be upper bounded by:
\begin{eqnarray}
    \sum_{i \in \C} \psi_{\sG, i} \le \frac{N-1}{N}\sum_{i=1}^N\mu_{i} =\sum_{i=1}^N\mu_{i}-{\bar{\mu}}.\label{gc_cond}
\end{eqnarray} 
Consider any NE $\bphi^*$ of $\P$ with $\bphi^*_{i^*} = \{a_1\} $. Then  by  \eqref{Eqn_monotone_utils_k_one}, the best response of $\C$ against $\bphi^*_{i^*} = \{a_1\}$   is,  for some $k \ge 1$ 
\begin{eqnarray}
    \bphi^*_{\sC}=\BR_{\sC}(\{a_1\})=\{\mbox{  $k$ times } a_1, a_2, \cdots,a_{N-k}\}, \nonumber  
\end{eqnarray}
(we will have $k\ge 1$,  if not, i.e., if  $a_1 \notin \bphi^*_{\sC}$, we have a contradiction to $\bphi^*_{\sC}$ being the best response, as by given hypothesis  $ \sum_{i=2}^N\mu_{i}<\sum_{i=2}^{N-1}\mu_{i}+\nicefrac{\mu_1}{2}$). 
By optimality of best response we have for any $l\ge 1$:
 \begin{eqnarray}\nu_{\sC}^\sP(\bphi^*) = U(\bphi_{\sC}^*,\bphi_{i^*}^*) &= &
    \frac{k}{k+1}\mu_1+\sum_{i=2}^{N-k}\mu_{i}  \nonumber \\ 
    &\ge& \frac{l}{l+1}\mu_1+\sum_{i=2}^{N-l}\mu_{i}  .  \hspace{1mm} \label{cond_NE1}
\end{eqnarray} 
Using the above with $l = 1$, and from 
 \eqref{int-cond} and \eqref{gc_cond}   we have:
\begin{eqnarray}
    \nu_{\sC}^\sP(\bphi^*)-\sum_{i \in \C} \psi_{\sG,i} 
    &\ge&\frac{\mu_1}{2}+ \sum_{i=2}^{N-1}\mu_{i}-\sum_{i=1}^N\mu_{i}+{\bar{\mu}}  \nonumber\\
&=& \bar{\mu}-\frac{\mu_1}{2}-\mu_{N}   \ 
> \ 0 . \label{cond_utility_1} 
\end{eqnarray}

Now consider the other case, i.e., any NE $\bphi^*$  with $\bphi^*_{i^*} = \{a\} $,  where $a \ne a_1$ (observe that $a \ne a_N$ either, as $\mu_N < \mu_1 /2$ contradicting optimality in the best response). The best response  against such $\bphi^*_{i^*}$  is, 
\begin{eqnarray}
    \bphi^*_{\sC}=\BR_{\sC}(\{a\})=\{a_1,\cdots,a_N\}\setminus \{a\} 
\end{eqnarray}
(if not, i.e., if $a \in \bphi^*_{\sC}$, we have a contradiction for $\bphi^*_{i^*}$ being the best response against $\bphi^*_{\sC}$ as then  $\nu_{\{i^*\}}^{\sP}(\bphi^*) =\nicefrac{\mu_a}{k}$ where $k = \gamma_a(\bphi^*)$ and for any $k \ge 2$, we have $ \nicefrac{\mu_a}{k} <\nicefrac{\mu_1}{2}$). Again by  the optimality of the  best response:
\begin{eqnarray} 
     \nu_{\sC}^{\sP}(\bphi^*)=\sum_{i=1}^N\mu_{i}-\mu_{a}&\ge& \sum_{i=1}^{N-1}\mu_{i}-\mu_{a}+\frac{\mu_{a}}{2}. \    \label{cond_2_utility}
 \end{eqnarray}
 Thus  we have   $ \mu_{N}\ge \nicefrac{\mu_{a}}{2}$ and further by hypothesis
 from   \eqref{gc_cond}: 
\begin{eqnarray}
    \nu_{\sC}^\sP(\bphi^*)-\sum_{i \in \C} \psi_{\sG,i} 
    \ge {\bar{\mu}}-\mu_{a}  
    \ \ge \ {\bar{\mu}}-2\mu_{N}   
    >  0  .\label{Diff_pos2}
\end{eqnarray} 

In all, from \eqref{cond_utility_1} and \eqref{Diff_pos2}, GC is blocked by $\C = \N-\{i^*\}$  under  pessimal anticipated utility as:
\begin{eqnarray*}
\hspace{19mm}    \underline{\nu}_{\sC}=\min_{\bphi^* \in \Phi^{\sP}} \nu_{\sC}^\sP(\bphi^*)> \sum_{i \in \C} \psi_{\sG,i}.   \hspace{10mm}  \mbox{\eop}
\end{eqnarray*} 

{\bf Proof of Part (ii):} 
Consider any partition $\P$ with size, $|\P|\ge k$ and $\beta = 0$ as in hypothesis.  We will  show below that 
there exists no NE with distinct s.
We are then done with the proof, as in view of Lemma \ref{lemma_2_unq_optimizer} the merger GC  blocks any P-NE pair with NE that has repeated s.

Say $\C \in \P$. Clearly cardinality $|\C|\le N-k+1$  
 and the equality is possible only when other $(|\P|-1)$ coalitions  are singletons and when $|\P|=k$.


 If possible  consider   NE $\bphi = \{a_1, \cdots, a_N\}$, i.e.,   NE with distinct s. We claim that $a_1$ and all the last $(N-k+1)$  s  $\{a_k, a_{k+1} \cdots, a_N\}$  are not in a single coalition -- as otherwise the cardinality of such a coalition would be greater than or equal to $(N-k+2)$. Say   $a' \in \bphi_{\sC}$ for some $\C \in \P$ and  $a' \in \{a_k, a_{k+1} \cdots, a_N\}$; and say $a_1 \notin \bphi_\sC$.  By hypothesis $\nicefrac{\mu_1}{2}>\mu_{a'}$,  and thus $\bphi_{\sC}$ is not  a best response, contradicting that $\bphi$ is an NE.  \eop

 \hide{
 Say, $a_1 \in \bphi_{\sC}$. Then there exists at least one  $a' \in \{a_k, a_{k+1} \cdots, a_N\}$ and $a' \notin \bphi_{\sC}$, because $|\bphi_{\sC}|=N-k+1$. Hence there exists a coalition $\C' \in \P$ such that $a' \in \bphi_{\sC'}$. 
 


Thus, for any NE  at $\beta=0$ the total utility of any partition $\P$ other than GC, is less than the worth of GC for the unique optimizer:
\begin{eqnarray}
    \sum_{\C \in\P} \sum_{i \in \C}\psi_i< \sum_{i=1}^N\mu_{i}={\nu}_{\sG}^{\sP_{\sG}}=\underline{\nu}_{\sG}
\end{eqnarray}

Thus, GC blocks any configuration involving $\P$ by merging and no partition with size $|\P|=k$ is stable as for all $\C \in \P$ (with size $|\P|=k$), the size of the coalition $|\C|\le N-k+1$, Thus for partitions with size $|\P|\ge k$ for all $1\le k\le N$ the argument is also true.\eop
}

{\bf Proof of Part (iii):} 
Consider any partition $\P$ with size, $|\P|> N-k$ and $\beta = 0$ as in hypothesis.  Similarly as the previous proof we will  show below that 
there exists no NE with distinct s.
We are then done with the proof. 

Say $\C \in \P$. Clearly cardinality $|\C|\le k$  
 and the equality is possible only when other $(|\P|-1)$ coalitions  are singletons and when $|\P|=N-k+1$.


 If possible  consider   NE $\bphi = \{a_1, \cdots, a_N\}$. We claim that $a_N$ and all the first $k$  s  $\{a_1, a_{2} \cdots, a_k\}$  are not in a single coalition as otherwise the cardinality of such a coalition would be greater than or equal to $k+1$. Say $a_N \in \bphi_{\sC}$; then there exists an    $a' \in \{a_1, a_{2} \cdots, a_k\}$ such that, $a' \notin \bphi_{\sC}$.  By hypothesis $\nicefrac{\mu_{a'}}{2}>\mu_N$,  and thus $\bphi_{\sC}$ is not  a best response, contradicting that $\bphi$ is an NE.  \eop

\RemoveCycle{ \begin{lemma}\label{lem_ne_exist}
  Strategy profile $\bphi  =   ((a_1, a_2, \cdots, a_{N-1}), a_1)$, where only $a_1$ is repeated two times is a   NE  for partition $\P=\{\C,\{1\}\}$   if,
     $   \mu_1<6\mu_{N-1}.$ 
 \end{lemma}
{\bf Proof:} By   given hypothesis  for any $k\ge 2$:
\begin{eqnarray}
    (k-1)\mu_{N-1}&>&\frac{k-1}{2(k+1)}\mu_1,   \mbox { and thus using \eqref{Eqn_monotone_utils_k_one}, }\nonumber\\
    \sum_{i=N-k+1}^{N-1}\mu_i &>& \frac{k-1}{2(k+1)}\mu_1,  \mbox{ and so, }\nonumber\\
    \frac{\mu_1}{2}+\sum_{i=2}^{N-1}\mu_i&>&\frac{k}{k+1}\mu_1+\sum_{i=2}^{N-k}\mu_i. \label{eqn_Ne_exist}
\end{eqnarray}
Then   the BR of $\C \in \P$, $\bphi_\C$ against $\bphi_{\{1\}}=  (a_1)$  is given by:
\begin{eqnarray}
    \BR_\C(a_1)=\{a_1, a_2 \cdots, a_{N-1}\}.
\end{eqnarray}
Again by \eqref{Eqn_monotone_utils_k_one} and the hypothesis the BR of the singleton player $\bphi_{\{1\}}$ against the BR of $\C$ is given by:
\begin{eqnarray}
    \BR_{\{1\}}(a_1, a_2 \cdots, a_{N-1})=\{a_1\}.
\end{eqnarray}
Thus, $\bphi=((a_1,a_2 \cdots, a_{N-1}),a_1)$ is an NE of partition $\P$. \eop
}

\end{document}